\documentclass[useAMS]{mn2e}
\usepackage{times}
\usepackage{psfig}
\usepackage{epsfig}
\usepackage{graphicx}

\def\lsim{\lower.5ex\hbox{$\; \buildrel < \over \sim \;$}}
\def\gsim{\lower.5ex\hbox{$\; \buildrel > \over \sim \;$}}

\def\lsim{\lower.5ex\hbox{$\; \buildrel < \over \sim \;$}}
\def\gsim{\lower.5ex\hbox{$\; \buildrel > \over \sim \;$}}

\def\be{\begin{equation}}

\def\ee{\end{equation}}

\begin{document}

\title[A white dwarf-neutron star relativistic binary model...]{A white 
dwarf-neutron star relativistic binary model for soft gamma-ray repeaters}

\author[Mosquera Cuesta]{Herman J. Mosquera Cuesta$^{1,2,3}$ \\
$^1$ Abdus Salam International Centre for Theoretical Physics,
Strada Costiera 11, Miramare 34014, Trieste, Italy \\ $^2$ Centro
Brasileiro de Pesquisas F\'{\i}sicas, Laborat\'orio de Cosmologia e
F\'{\i}sica Experimental de Altas Energias \\ \hskip 0.2 truecm 
Rua Dr. Xavier Sigaud 150,
Cep 22290-180, Urca, Rio de Janeiro, RJ, Brazil \\ $^3$ Centro
Latino-Americano de F\'{\i}sica, Avenida Wenceslau Braz 173, Cep
22290-140 Fundos, Botafogo, Rio de Janeiro, RJ, Brazil }

\maketitle

\date{\today}

\begin{abstract}
A scenario for SGRs is introduced in which gravitational radiation
reaction effects drive the dynamics of an ultrashort orbital period
X-ray binary embracing a high-mass donor white dwarf (WD) to a rapidly
rotating low magnetised massive neutron star (NS) surrounded by a
thick, dense and massive accretion torus. Driven by GR reaction,
sparsely, the binary separation reduces, the WD overflows its Roche
lobe and the mass transfer drives unstable the accretion disk around
the NS. As the binary circular orbital period is a multiple integer
number ($m$) of the period of the WD fundamental mode (Pons et al.
2002), the WD is since long pulsating at its fundamental mode; and most
of its harmonics, due to the tidal interaction with its NS orbital
companion. Hence, when the powerful irradiation glows onto the WD; from
the fireball ejected as part of the disk matter slumps onto the NS, it
is partially absorbed. This huge energy excites other WD radial ($p$-mode)
pulsations (Podsiadlowski 1991,1995). After each mass-transfer episode
the binary separation (and orbital period) is augmented significantly
(Deloye \& Bildsten 2003; Al\'ecyan \& Morsink 2004) due to the
binary's angular momentum redistribution. Thus a new adiabatic inspiral
phase driven by GR reaction starts which brings the binary close again,
and the process repeats. This model allows to explain most of SGRs
observational features: their recurrent activity, energetics of giant
superoutbursts and quiescent stages, and particularly the intriguing
subpulses discovered by BeppoSAX (Feroci et al. 1999), which are
suggested here to be {\it overtones} of the WD radial fundamental mode
(see the accompanying paper: Mosquera Cuesta 2004b). 
\end{abstract}



\section{Astrophysical Motivation}

The giant outburst from SGR1900+14 on 27 August (1998), hereafter
GRB980827, was an extraordinary event in many respects. Aside from the
huge energy released ($E_{out} \geq 3\times 10^{44}$~erg) on such a
short timescale ($\sim 0.1$~s), the clear identification in its light
curve (and power spectral density) of a main pulsation period ($P \sim
5.16$~s) and an accurately measure interpulse set ($\sim 1$~s of
interval), as discovered by BeppoSAX (Feroci et al.  1999), has
challenged the current views on SGRs. This behavior, Feroci et al.
(1999) advanced, is unexpected and quite difficult to explain in the
magnetar framework. The  GRB980827 subpulses seem to be {\it
overtones} of the fundamental frequency $f_0 = 0.194$~Hz (Feroci et al.
1999). As shown by Mosquera Cuesta  (2004b), where a consistent
scenario to explain the observed SGR1900+14 modulations was presented,
when looking at the GRB980827 lightcurve perhaps we are seeing the full
WD pulsation spectrum. In these lines, it would be crucial if the {\it
propeller model} by Marsden, Rothschild \& Lingenfelter (2001) and
Alpar (1999,2000) could provide a suitable explanation for the
pulsational modes observed in GRB980827, since this way we can
discriminate among accretion models.  

The rapid risetime (0.25~s), energetics and subpulses seen by BeppoSAX 
in this powerful event are characteristics challenging the magnetar 
view. According to the magnetar scenario (Thompson \& Duncan 1995,1996) 
the typical (crustal stored stress) energy of an outburst from an SGR
source should be $\sim 10^{39}$~erg (de Freitas Pacheco 1998) if one
expects the source not to be destroyed by the explosion. Energies as
$E_{out}$ or higher can be obtained at the expense of destructing
practically the whole crust of the NS undergoing the starquake (de
Freitas Pacheco 1998) or through magnetic reconnection (Thompson \&
Duncan 1995,1996; Woods et al. 1999).  However, a NS with no crust is 
fully unstable, and should explode (Colpi, Shapiro \& Teukolsky 1993).
Consequently, the source could not repeat anymore.

The problems for the magnetar model do not stop there. After a burster
phase in 1996, Marsden, Rothschild \& Lingenfelter (1999, see also
Harding, Contopoulos \& Kazanas 1999; Murakami et al. 1999) found that
the pulsar doubled its period derivative, $\dot{P}$, which would
suggest the pulsar magnetic field energy had augmented $\sim 100\%$
during the outburst, what clearly opposes the magnetar scenario. They
concluded that the SGRs spindown is due not to dipole radiation but to
another mechanism. Thus, the $\dot{P}$ could not be an estimate of the
magnetic field strength, nor evidence for a magnetar. More recently,
Rothschild, Marsden \& Lingenfelter (2001) argued that the unexpected
occurrence of most of the SGRs (and AXPs) in the denser phases of the
interstellar medium effectively {\it rules out} the magnetar model for
their origin, since in that model the NS endowes a supercritical field
as an intrinsic property, with no plausible relation to external
environments. In addition to those shortcomings of the magnetar scenario, 
very recently P\'erez Mart\'{\i}nez et al. (2003) rose questions on the
formation of the magnetar itself by invoking quantum-mechanical effects
on the structural stability of the hypermagnetised star, while Rezzolla 
\& Akhmedov (2004) demonstrated that a fundamental change in the Pacini's
spindown law appears when describing the electromagnetic fields in full
general relativity.

Then the issue is contentious: what are these objects?

In this letter we suggest that SGRs may not be related to magnetars, 
as opposed to claims by Kouveliotou et al. (1998,1999); Hurley et al.
(1999a,b,c); Murakami et al. (1999); Mazets et al. (1999a,b); etc.
Instead, it is shown that SGRs may be extremely close (tight)
relativistic X-ray binaries with periods near {\it a few tens of
seconds} (see Table.1)\footnote{It is worth quoting that Kouveliotou's 
team (1999)
found (0.05 cycles) systematic departures of pulse phases from the
best-fit ephemeris. Due to its potentiality, searches for an orbital
period in the SGR 1900+14 August, 1998 observations were performed.
They divided the data set in subsets of 400 seconds, and conducted
searches in the range:  $10^3 - 10^6$~s.  No sinusoidal modulation 
was found.  If our model is on the right track, the negative result 
is not surprising for a shorter period for the SGR 1900+14 binary, 
$21.4 \leq P_{\rm SGRs} \leq 65$~s, but expected. Thus, it is 
likely that the orbital signature looked for remains embedded in the 
data subsets that were used for.}, in which sparsely the rate of 
shrinking of the binary separation dominates over the massive WD rate 
of contraction below its equilibrium radius (Fryer et al. 1998b), and 
mass transfer ensues. To trigger a superoutburst the system is brought
closer again due to GR emission. In a timescale $\Delta T_{rep}\leq
10$~yr the process described above restarts governing the binary
dynamics leading to a new outburst.

\section{SGRs as WD-NS Relativistic Binaries}

\subsection{Origin and evolution driven by GR reaction}

According to King \& Ritter (1998) an interesting by-product of the
massive case-B evolution model of Cygnus X-2 is that binaries
constituted by a massive WD ($M_{WD} \geq 1.0$~M$_\odot$) and massive
rapidly rotating ({\it millisecond}) accretion spun up NS ($M_{NS} =
1.78 \pm 0.23$~M$_\odot$), with ultrashort orbital periods ($0.004 \leq
P_f (d) \leq 0.02$) can exist for very massive WD companions.  Systems
with even more shorter, near tens of a second, depending on the common
envelope efficiency parameter $\alpha_{CE}$, can be formed (see also
Podsiadlowski 1995,2000; Ergma, Lundgren \& Cordes 1997; Tauris, van
den Heuvel \& Savonije 2000). This view gives us some insight on the
SGRs former evolution. Perhaps those systems come from a previous
Thorne-Zytkow object in which a complete exhaustion or ejection of the
hydrogen-rich envelope of the red giant star (lifetime $\tau_{RG} \sim
10^4$~yr) occurred once the NS is engulfed in a common-envelope
evolutionary phase caused by shrinking of the orbit due to
gravitational-wave (GW) radiation-reaction effects. This process leaves
the giant companion's bare Helium (or CNO) core, the WD, in a tight
orbit with a rapidly spinning massive NS. We suggest this may explain
why SGRs are surrounded by nebulae resembling supernovae remnants. The
NS high mass being a consequence of the Thorne-Zytkow object latest
stage, while the NS spin stems from its former LMXB evolution (van
Paradijs 1995a,b). A similar evolutionary path was described by Ergma,
Lundgren \& Cordes (1997), see also Tauris, van den Heuvel \& Savonije
(2000). Ergma et al. showed that these systems may undergo a second
mass transfer transient around periods of 5 minutes or less, being the
source observed as an X-ray binary again. The final fate could be the
binary coalescence. We also note that in the last few years a large
number of systems of this sort have been discovered (Edwards \& Bailes
2001; Tauris \& Sennels 2000; Tauris, van den Heuvel \& Savonije  2000,
a large sample is collected in this reference; Deloye \& Bildsten 2003,
Al\'ecyan \& Morsink 2004), and their formation have been modelled with 
great details (Ergma, Lundgren \& Cordes 1997; Tauris, van den Heuvel \& 
Savonije  2000). At this late stages GW dominate the binary dynamics and 
secularly shorten its period to those we suggest here.

\begin{table}
\centering
\begin{minipage}{80mm}
\caption{A set of viable orbital periods, GR timescale and binary separation 
for the ultracompact relativistic WD-NS binary. The minimum period compatible 
with all constraints from SGR1900+14 is $P_{\rm min} \sim 21.8$~s.\label{tbl-0} }
\begin{tabular}{ccc}
\hline
{ Period } & { GR-Time \footnote{Computed using Peters \& Mathews relation} } 
 &  { Orb. Sep. \footnote{Computed using  Kepler's third Law} }   \\
{ s } &        { yr }  &         {  km }      \\
\hline
  20.0   &   3.418   &    8252.0    \\
  22.0   &   4.407   &    8794.0    \\
  24.0   &   5.558   &    9319.0   \\
  26.0    &  6.881   &    9830.0    \\
  28.0   &   8.384   &   10330.0    \\
  30.0   &   10.08   &   10810.0    \\
  32.0   &   11.97   &   11290.0    \\
  34.0   &   14.07   &   11750.0     \\
  36.0   &   16.39    &  12210.0   \\
  38.0    &  18.93   &   12660.0   \\
  40.0   &   21.70   &   13100.0    \\
  42.0   &   24.72   &   13530.0   \\
  44.0   &   27.98   &   13960.0     \\ 
  46.0   &   31.51   &   14380.0    \\
  48.0   &   35.29   &   14790.0    \\
  50.0   &   39.35   &   15200.0   \\
  52.0   &   43.69   &   15600.0  \\
  54.0   &   48.32   &   16000.0    \\
  56.0   &   53.24   &   16390.0   \\
  58.0   &   58.46   &   16780.0    \\
  60.0   &   63.99   &   17170.0     \\
\hline
\end{tabular}
\end{minipage}
\end{table}


\subsection{SGRs as interacting binaries}

By the time the WD-NS relativistic binary reaches the critical
separation {\it transient} mass transfer occurs.\footnote{ We address
the readers to a couple of papers by Deloye \& Bildsten (2003) and
Al\'ecyan \& Morsink (2004); and references therein, where a similar
and detailed study of the dynamics of these kind of ultracompact,
ultra-relativistic binary systems driven by gravitational radiation
reaction was presented. Our model here can be thought of then as a
restricted case of the more general interaction they considered.
However, we alert that the WD in their study cases is a very low mass
companion. The implications of this feature for the minimum periods
that the binary can achieve (much less than three minutes in our
proposal, see Table.\ref{tbl-0}) cannot be overlooked. In overall, a
fundamental characteristic of these interacting binaries is that after
each mass-transfer episode the binary separation (and orbital period,
consequently) is augmented significantly, largely for non-circular
orbits. Then, after a new adiabatic spiral-in phase
driven by GR reaction the binary is brought close again, and the
process repeats.  This feature is a very key ingredient of our
scenario.} In a rather catastrophic episode, which repeats sparsely
any:  $\Delta T_{\rm SGRs} \leq 10$~yr (estimated below), the WD starts
to transfer mass onto a low-magnetised rapidly rotating massive
(2~M$_\odot$) NS, via the formation of a thick dense massive accretion
disk (TDD) very close to the innermost stable circular orbit (ISCO).
The disk becomes unstable due to gravitational runaway or Jeans
instability, partially slumps and inspirals onto the NS. The abrupt
supercritical mass accretion onto the NS releases quasi-thermal
powerful $\gamma$-rays, a fireball to say, while triggers
non-radial NS oscillations which emit fluid-mode GW (see Mosquera
Cuesta et al. 1998). A parcel of the accretion energy illuminates with
hard radiation the WD perturbing its hydrostatic equilibrium. The WD
absorbs this huge energy at its interior and atmosphere and begins to
radially pulsate ($p$-modes) in addition to its long before excited
fundamental mode and higher harmonics ({\it overtones}: Podsiadlowski 
1991,1995), see Mosquera Cuesta (2004b), and Tables \ref{tbl-1}-\ref{tbl-2}, 
below.

The accretion rate from the WD onto the TDD, to be replenished during
the periastron passage $\Delta T_{p-a}$ around the NS, can roughly be
estimated from an idealized {\it drop model} in which we assume that
the total mass contained in a sphere of radius equivalent to half the
WD tidal height $H_{tide}$ (Shapiro \& Teukolsky  1983) drops onto the 
NS: $ H_{tide} = R_{WD}\left(\left[\frac{M_{NS}}{M_{WD}}\right]^{1/3} 
- 1\right) \leq 1100 {\rm \; km}$. Using this, and a mean density 
$\bar{\rho}_{WD} = 4\times 10^6 {\rm g~cm^{-3}}$, we obtain the WD mass 
transfer rate

\be 
\dot{M}{^{WD}_{TDD}} \sim \left[\frac{M_{Lost}^{WD}} {6\times 10^{-4} {\rm 
M_\odot}}\right] \left[\frac{ 10 \rm s }{\Delta T_{p-a}}\right] \sim 6 \times 
10^{-5} {\rm M_\odot ~ s^{-1}},
\ee

where we have used results for $R_{WD} \sim 4.7\times 10^3$~km by Suh \& 
Mathews (2000) and Eggleton's theory (1983), see below.

Since essentially viscosity dominates the TDD hydrodynamics, in general one
can estimate the viscous timescale (Frank, King \& Raine 1992) for it to 
drive unstable the TDD as (Popham, Woosley \& Fryer 1999)

\begin{eqnarray}
\Delta T_{visc} \sim \frac{R^2}{\nu} =  \left[\frac{R^2}{H^2 \alpha 
\Omega_{\rm K}}\right] \sim 4 \times 10^{-3} \frac{ M^{-1/2}_{NS} 
R_{6}^{3/2} }{ \alpha_{0.1} } {~\rm s}, \label{viscosity}
\end{eqnarray}

with $\alpha$, $H$ and $\Omega_{\rm K}$ the Shakura-Sunyaev parameter,
disc height scale and Keplerian angular frequency, respectively. Thus a
rough estimate of the mean NS accretion rate from the TDD reads

\begin{eqnarray}
\dot{M}{_{NS}^{TDD}} \sim  \frac{M{^{WD}_{TDD}}}{\Delta T_{visc}} = 0.37 
\alpha_{0.1} M{^{WD}_{TDD}} M_{NS}^{1/2} R_{6}^{-3/2}\; ,
\end{eqnarray}

where $\dot{M}_{NS}^{TDD} \sim 2 \times 10^{-5} {\rm M_\odot}~{\rm s^{-1}}$ 
is the mass deposited at radius $R_6$ (in units of $10^6$~cm). Thus, we 
conjecture that giant outbursts from SGRs, like GRB980827, are triggered when 
a large part of the accretion torus plunges onto the NS.

Scaling these results for the difference in gravitational
potential between a BH and a NS ($\Phi_{NS}/\Phi_{BH} \sim 0.2)$,
in coalescence with a WD (Klu\'zniak \& Lee 1998; Klu\'zniak 1998),
and using  the Eq.(1) given by Mosquera Cuesta et al. (1998) 
to estimate the peak temperature achieved when the 
accreted matter crashes on the NS surface, we get  

\begin{eqnarray} 
T_{peak}  =  \left[\frac{G \beta_{acc}}{4\pi\sigma_{SB}}\right] 
\left[\frac{ M_{NS} \dot{M}{_{NS}^{TDD}} }{ R{_{NS}^3} }\right]^{1/4} 
= 3.1\times 10^{10} {\rm K},
\end{eqnarray}

or equivalently, $T_{peak} = 2.1$~MeV, and a total energy released
during this transient is $E_{Burst} >> 10^{44}$~erg. Both results match
very nicely the power emitted in GRB980827 (Mazets et al. 1999; Feroci
et al. 1999). It also gets close to total energy that can be released
in GRBs from galactic sources, as discussed by Fryer et al. (1998b).

\begin{table*}
\centering
\begin{minipage}{140mm}
\caption{SGRs observed properties and  WD viscosity and mass (assuming 
$\Gamma = 5/3$) according to our binary scenario.\label{tbl-1} } 
\begin{tabular}{cccccccc}
\hline
{Source} & { Pulsation } & {Luminosity} & {Released} & {Temperature} & 
{Distance} & {Viscosity} 
& {WD  Mass}  \\
{  } & {Period (s)} & { (ergs$^{-1}$) } & {Energy (erg)} & {Peak (MeV)} &  
{(kpc)} & {Mean ($<\nu>_{13}$)} & {(M$_\odot$)} \\
\hline
{SGR 1900+14} & 5.16 & { $3 \times 10^{41}$} & $3.4 \times 10^{44}$ & 1.2 & 
{ 5.7 } & { 1.0 } &  1.1  \\
{SGR 0526-66} & 8.1 & { $6 \times 10^{41}$} & $ 6\times 10^{45}$ & 2.1 & { 55 } 
& { .... } & 0.70  \\
{SGR 1806-20} & 7.47 & { $4 \times 10^{41}$} & $ 3\times 10^{42}$ & $ 0.3 $ & 
{ 14  } & { .... } & 0.80 \\
{SGR 1627-41} & 6.7 & { $8\times 10^{43}$} & $3.0 \times 10^{42}$ &  {1.0} & 
{ 5.8 } & { .... } & {0.95 } \\
\hline
\end{tabular}
\end{minipage}
\end{table*}

\subsection{A possible signature of a WD in SGR1900+14}

The WD atmospheric temperature after being flared should
be far much higher than the one for surface thermalization (Ergma,
Lundgren \& Cordes 1997) 

\be
T = \left(\frac{L|{ ^{\rm SGR1900+14}_{fireball} } }{4 \pi R{^2_{WD}} 
\sigma_{SB} } \right)^{1/4} = 6.8 \times 10^7~ {\rm K},
\ee

where 

\be
L|{ ^{\rm SGR1900+14}_{fireball} } \equiv G M_{NS}
\dot{M}{_{NS}^{TDD}}/R{^{TDD}_{NS}} \sim 10^{43-44}~ \rm erg \; s^{-1}\;, 
\ee

is the accretion luminosity, as given above. Therefore, the star emits
X-rays relatively hard, and just in the band width of Rossi XTE,
BeppoSAX, CHANDRA, XMM-NEWTON, INTEGRAL, etc. By the time its
atmospheric temperature rolls down the Debye temperature $\Theta_D \sim
10^7$~K due to reemission, the sudden {\it crust} crystallization
increases the WD cooling rate because the surface specific heat is now
due to {\it lattice vibrations} instead of thermal motions (Shapiro \&
Teukolsky 1983).  Assuming $\sim 10\%$ of the trapped energy is quickly
returned into space, the timescale for this transition to occur is (see
Table 1 and Figure 1 of Feroci et al.  1999)

\be
\tau_{cryst} \sim \frac{ 0.1 E_{Burst} } { L|{^{\rm SGR1900+14}_{fireball} } }  
\sim 100 ~{\rm s}.
\ee

Then our view predicts: the very noticeable transition of the
exponential {\it time constant} in GRB980827 lightcurve, around $\sim
80$s after risetime, could be a signature of a crystallization phase
transition of the WD crust. We also stress that by using
Eq.(\ref{viscosity}) some insight on the WD viscosity, over the event
duration, could be inferred from a careful analysis of GRB980827
lightcurve of $\sim 300$~s. In this respect, a critical analysis of this
lightcurve shows no evidence of a sinusoidal modulation; the signature of 
ocultation in the binary. Hence, the line-of-sight angle must satisfy 
$i \geq 45^o$ for the system orbital features.

\subsection{GWs at action}

Assuming a polytropic model for the WD with $\Gamma = 5/3$,
the mass-radius relation is given by (Eggleton 1983)

\begin{eqnarray}
R_{WD} \simeq  10^4 {\left[\frac{M_{WD}}{ 0.7 \; M_\odot}\right]}^{-1/3}
{\left[1 -\left(\frac{M_{WD}}{M_{Ch}}\right)^{4/3}\right]}^{1/2}
{\left[\frac{\mu_e}{2}\right]}^{-5/3},
\end{eqnarray}

where $R_{WD}  =  4.8 \times 10^3 \; {\rm km}$ is the WD radius,
$M_{WD} \sim 1.1$~M$_\odot$ the WD mass (justified in the related paper
by Mosquera Cuesta 2004b), $M_{Ch} = 1.44$~M$_\odot$ (Chandrasekhar
limit), and $\mu_e \sim 2$ the molecular weight per electron. The
orbital separation for which mass overflow commences is given by
(Eggleton 1983)

\be 
a_0 = R_{WD}\left[{0.6 q^{2/3} + \ln (1 + q^{1/3})}/{0.49 q^{2/3}}\right],
\ee

with $q \equiv M_{WD} / M_{NS}$ the mass ratio. Thus, we obtain $a_0 = 8.8
\times 10^3$~km. 

Since the criterion for stable mass transfer is satisfied ($q\leq
2/3$), as the WD loses mass its orbit should widen to replace it just
below its critical Roche lobe separation. This can be accomplished
through the formation of the TDD around the NS in a timescale shorter
than the periastron time

\be
\Delta T_{TDD} \sim  2 \times \frac{ R^{WD}_{\rm Orb} }{ V^{WD}_{\rm Orb} } 
\sim \frac{8895~ \rm km} {5\times 10^3 \rm ~km~s^{-1}} \sim 1.8~ {\rm s}\; , 
\ee

which is shorther then the ``periastron'' passage time $\Delta T_{p-a}
\sim 6$~s. The angular momentum lost by the WD (transferred to the TDD
`stably' orbiting the NS) is returned back to the orbit due to angular
momentum conservation.  Including angular momentum losses to the just
formed accretion disk, the binary orbital separation by the time mass
transfer starts is given by (Podsiadlowski 1995)


\begin{eqnarray}
a & = & a_0 \left[\frac{M_{WD} + M_{NS+TDD} }{M{^0_{WD} } +
M{^0_{NS}}}\right] \left(\frac{M_{WD} }{M{^0_{WD}} } \right)^{C_1}
\left[\frac{M_{NS}}{M{^0_{NS}}} \right]^{C_2} \nonumber \\
& = & (1 + \bar{d}) a_0,
\end{eqnarray}

where $\bar{d} \leq 1$ (which means that the orbit can even double its
size in this case),  the superscript $^0$ stands for the states before
Roche lobe overflowing and the values for the constants $C_1$ and $C_2$
are: $C_1 \equiv -2 + 2J_{TDD} B, \hskip 0.1 truecm C_2 \equiv -2 -
2J_{TDD}$, with $B \sim 7\times 10^{-5}$ the fraction of the WD mass
that goes to the TDD formation. Furthermore, $M_{NS+TDD} \equiv B
(M{^0_{WD}} - M_{WD}) + M{^0_{NS}}$, and  no angular momentum is loss
from the binary $J_{eject} = 0$.

Assuming that the WD orbit is circular ($e \sim 0$: Pons et al. 2002)
and the orbital period today is near the one for mass transfer to start
at $a_0$, i.e. $P \sim 21$s\footnote{Note that an increase in orbital
period by a factor of 3 (still consistent with our picture) would yield
a time scale $\tau_{GRR} \sim 8000$~yr for the binary to coalesce
driven by GW.}, the timescale for the orbit to shrink due to effects of
radiation of GWs is given as


\be
\tau_{GRR} = \left(\frac{ a^4_0}{ 4\beta}\right) \leq 100~{\rm yr},
\ee

where 

\be
\beta \equiv ({64 G^3}/{5 c^5}) M_{WD} M_{NS} (M_{WD} + M_{NS})\; , 
\ee

with the GWs orbital angular momentum loss by the system

\be
{({dJ}/{dt})_{GW}} = {[{32 G^{7/2} }/{5c^5}]} {(M^{1/2}[{ M{^2_{WD}} 
M{^2_{NS}} }/{a^{7/2}}])}\; .
\ee

It is easy to see that if one wants the  SGR binary to reenter a new
mass transfer transient, a very short reduction in the binary
separation is required. Distance separation reduction (driven by GR
reaction effects) of {\it a few} hundred km will yield a timescale
compatible with the mean one observed for rebursting in most of the
SGRs, i. e., $\Delta T_{rep} \leq 10$yr. Since after mass-shedding the
orbit widens faster than the WD expands (i.e., WD recoils due to
angular momentum conservation [gravothermal effect] and stops mass
transfer), the angular momentum redistribution will prevent the binary
merger to occur on this timescale $\tau_{GRR}$. Consequently we can
expect the system to repeat several superoutbursts before the WD
thermal runaway final explosion or tidal disruption.  We also highlight
that the action of orbital gravitational stresses (powerful tides
driven by the NS) may drive a sort of plate tectonics on the WD crust,
{\it \`a la} Rothschild, Marsden \& Lingenfelter (2001), which can
provide enough energy so as to explain why SGRs glow in hard X-rays
over some months before undergoing dramatic transients such as
GRB980827 from SGR 1900+14.  This prospective mechanism will be 
tackled elsewhere.

\begin{table*}
\centering
\begin{minipage}{140mm}
\caption{Theoretical pulsation modes of a WD with $\sim$ 1.1 M$_\odot$, 
(as computed by Montgomery \& Winget 1999) versus the modulation spectrum 
from SGR 1900+14 on August 27 (1998) discovered by BeppoSAX. Also, SGRs 
pulsation periods and expected masses for the WD in the binary model.
\label{tbl-2} }
\begin{tabular}{cccc|ccc}
\hline
{S.H.R.Overtones.} &  {Num. Model [Hz]} & {BeppoSAX [Hz]} & {Mismatch (\%)} & 
{SGR} & {P [s]} & {Mass [M$_\odot$] } \\
\hline
{l= 0, n = 1} & { 0.1846} & {0.194} & { 5.0} & {1900+14} & 5.16 &  { 1.1 }  \\
{l= 1, n = 1} & { 0.3898} & {0.389} & {0.2} & {0526-66} & 8.1 & { 0.90 }  \\
{l= 2,0, n = 2,3} & {0.7644, 0.7845} & { 0.775} & {1.0, 1.23} & {1806-20} & 7.47 
& 0.95  \\
{l= 1,2, n = 3,3} & {0.9025, 1.0068} & {0.969} & {4.54, 3.78} & {1627-41} & 6.7 
& {1.0 } \\
{l= 1, n = 4} & {1.1430} & { 1.161 } & {1.55} \\
\hline
\end{tabular}
\end{minipage}
\end{table*}


\begin{figure}
{\centering \hskip 1.25 truecm \epsfxsize=6.6cm \epsfbox{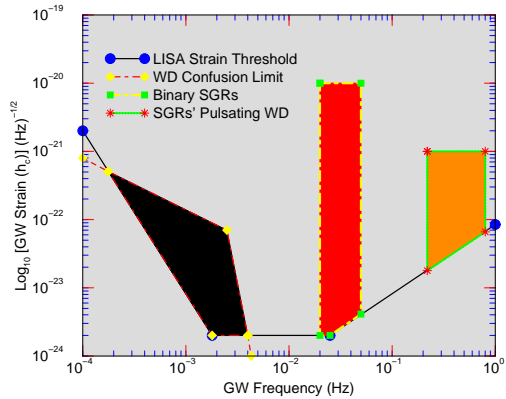} }
\caption{Prospectives for detectability by LISA of SGRs and their WD components 
in the relativistic binary model.}
\end{figure}

\section{SGRs GWs emission: A pathway to unravel their nature}

Several GW signals of different nature are expected to be produced in the
context of this picture for SGRs. During: a) the binary inspiral
(detectable by the LISA antenna) with characteristic GW amplitude

\be
\rm h_c(f) = 3.2\times 10^{-19} \left[\frac{\mu}{1 M_\odot}\right]^{5/3} 
\left[\frac{f}{ 3\times 10^{-2} Hz}\right]^{2/3} \left[\frac{5.7 kpc}{D}\right].
\ee 

b) the inspiraling of a lump of matter that reaches the disk at a radius 
$R \sim 100$~km; the equilibrium distance from the NS, c) the plunging of
the inner TDD, after crossing the ISCO, onto the millisecond NS. This will
shake the NS and its ({\it fluid-modes}), d) WD non-radial pulsations, and 
e) the NS spinning or wobbling.  Some of them having characteristics so as 
to make it detectable even with today's bars.  For the sake of definiteness, 
we compute here the characteristics of the GW burst released during the brief
inspiraling (timescale $\Delta T_{visc} \equiv \Delta t$) of a disk ``blob" 
till finding the ISCO.  The GW amplitude reads (Mosquera Cuesta et al. 1998)

\be
{\left|\frac{\Delta h}{\Delta t}\right|}^2 = \frac{4 G} {{c^3} }
\left(\frac{1}{D^2}\right) \frac{\Delta E_{GW}}{\Delta t},
\ee

with the GW energy $\Delta E_{GW} \equiv \Delta L_{GW} \times \Delta T_{visc}$.  
The GW luminosity 

\be
\Delta L_{GW} \sim (G/5c^5) \left[\dot{M}^2{^{WD}_{TDD}} \; R_{giro}^4 \; 
\Delta T_{visc}^{-4}\right]\; , 
\ee

where $R_{gyro} = 7\times 10^6$~cm is the radius of gyration of the
matter in the disk.  This yields $h_c = 1.5 \times 10^{-22}$ for a
source distance of 5.7 kpc (Kouveliotou et al. 1999).  The GW frequency
is given by $f_{\rm GW} \sim 2 \times (\Delta T_{visc})^{-1} \sim 480
\;{\rm Hz}$.  A GW signal such as this from SGR 1900+14 during the
GRB980827 energetic outburst could have been detected by
interferometers such as LIGO, VIRGO and GEO-600 were they operatives,
but may be detected in the future.  In passing, it worths to quote that
since the conditions for GWs resonant excitation (Ferrari et al. 2000)
of the stars' fluid modes (mainly the WD, $P_{orb} \sim 4 \times f_0$)
a continuous GWs emission could also be enhanced in this extremely
close binary.  This issue will be addressed elsewhere.

\section{Discussion and conclusion}

An observationally consistent scenario to explain the detected
modulations from SGR 1900+14 is discussed in the accompanying paper by
Mosquera Cuesta (2004b). In that paper is also provided a viable 
origin for the secular SGRs spindown, and for the up-and-down of their
periods. Most of the results are collected in Table 2. Related to those
results, we stress that neither NS-NS nor NS-BH binaries can fit the
observations of SGRs (Kouveliotou 1998,1999; Hurley 1999a,b,c; Murakami
et al. 1999; Feroci et al. 1999), since, even for the explosive lower
mass limit of Colpi, Shapiro \& Teukolsky (1993), $M_{NS} \leq 0.09$~M
$_\odot$, the oscillation frequency of its fundamental mode goes too
high (McDermott, van Horn \& Hansen 1987) due to the NS higher density,
so as to explain the relative lower pulsation periods observed in SGRs
(see Table\ref{tbl-1}).  Moreover, since the maximum mass of a WD is 
limited to the Chandrasekhar mass: $M_{Ch} = 1.44$~M$_\odot$, we expect 
not to find SGRs with pulsation periods $\leq 2.1$~s (see Table\ref{tbl-1}). 
A prediction that agrees with current data, but that is not precluded by 
the magnetar model.

To conclude, the NS low magnetization in this model places it under the
pulsar death line in the $P-\dot{P}$ diagram, and hence turns SGRs
undetectable as binary radio pulsars; a point confirmed by Xilouris et
al. (1998).  This fact makes the search for an optical and/or infra-red
counterpart of SGRs with GEMINI or KECK a timely endeavour. Meanwhile,
if our scenario proves correct, a prospective way to study these
systems is through the new generation of X-rays telescopes like
CHANDRA, XMM-Newton, INTEGRAL, etc., and GWs observatories such as
LIGO, VIRGO, TAMA-300, GEO-600, the TIGAs Network (which may follow-up
the burster phase) and LISA (which may follow-up the orbital dynamics
and WD pulsations). When operatives, they will play a decisive role in
deciphering the SGRs nature, as perhaps, new general relativistic
astrophysical laboratories where WD-NS interacts farther out the 
weak-field regime.

I thank Professors Jhon C. Miller (Oxford U.-SISSA), Jos\'e A. de Freitas
Pacheco (OCA-Nice), Jorge E. Horvath (IAG-Sao Paulo) and Luciano Rezzolla
(SISSA) for fruitful discussions and insight during the early days of
the development of this idea.


\begin{thebibliography}{99}
\frenchspacing

\bibitem[]{morsink04}Al\'ecyan, E. and Morsink, S. M., 2004,  ApJ in press

\bibitem[]{alpar99}Alpar, M. A., 1999, report astro-ph/991228; 2000. astro-ph/0005211,

\bibitem[]{colpi93}Colpi, M., Shapiro, S. L. \& Teukolsky, S. A., 1993, 
ApJ 414, 717

\bibitem[]{pacheco98}de Freitas Pacheco, J. A., 1998, Astron. \& Astrophys., 
{336}, 397. See also  Blaes, et al., 1989, ApJ, {343}, 839

\bibitem[]{Deloye}Deloye, C. J., and Bildsten, L., 2003, ApJ 598, 1217

\bibitem[]{DT92}Duncan, R. C. \& Thompson, C., 1992, ApJ, {392}, L9

\bibitem[]{duncan98}Duncan,  R. C., 1998, ApJ, {498}, L45

\bibitem[]{edwards}Edwards, R. T. \& Bailes, M., ApJ 547, L37.


\bibitem[]{ergma97}Ergma, E., Lundgren, S. C. \& Cordes, J., 1997, ApJ, {475}, 
L29

\bibitem[]{feroci99}Feroci, M., et al., 1999, ApJ, 515, L9

\bibitem[]{ferrari00}Ferrari, V., et al., 2000, Int. J. Mod. Phys. D 9, 495 

\bibitem[]{frank92}Frank, J., King, A. R. \& Raine, A, 1992, Accretion power 
in astrophysics, CUP Press, Cambridge, UK 


\bibitem[]{fryer98-2}Fryer,  C. L. \& S. E. Woosley, 1998a, ApJ, {\bf
502}, L9

---- Fryer, C. L., et al., (1998b) report astro-ph/9808094 

\bibitem[]{harding99}Harding, A., Contopoulos, I \& Kazanas, D., 1999, Ap. 
J. 525, L125,

\bibitem[]{hurley99-2}Hurley, K., et al., 1999a, ApJ, {510}, L107

---- {\it Ibid.}, 1999b, {510}, L111

---- {Nature}, 1999c, {397}, 41

\bibitem[]{king98}King, A. R. \& Ritter, H., 1998, report astro-ph/9812343

---- 1997, ApJ, {482}, 919


\bibitem[]{kluzniak98}Klu\'zniak, W. \& Lee, W., 1998, ApJ,
{494}, L53

\bibitem[]{kluzniak98a}Klu\'zniak, W. 1998, ApJ, {509}, L37


\bibitem[]{kouveliotou99-2}Kouveliotou, C., et al., 1998, Nature, {393}, 235

---- 1999, ApJ, {510}, L115

\bibitem[]{McDVanHan}MacDermott, P. N., Van Horn, H. M. \&  Hansen, C. J.,
ApJ, {325} 725 (1989).



\bibitem[]{marsden99}Marsden, D., Rothschild, R. E. \& Lingenfelter, R. E.,
1999, ApJ, 520, L107




\bibitem[]{mazets99}Mazets, E., et al., 1999, report astro-ph/9905195 and 
astro-ph/9905196

\bibitem[]{montgomery}Montgomery, M. \& Winget, D. E., 1999, ApJ
 526, 976



\bibitem[]{herman}Mosquera Cuesta, H. J., et al., 1998, Phys. Rev. Lett.,
{80},  2988

\bibitem[]{herman99-2}Mosquera Cuesta, H. J., de Ara\'ujo, J. C. N., Aguiar, 
O. D. \&  Horvath, J. E., 2000, CBPF-NF-083, 7pg

\bibitem[]{herman01}Mosquera Cuesta, H. J., 2004b, accompanying paper submitted


\bibitem[]{murakami99}Murakami, T., et al., 1999, ApJ,
{510}, {L119}

\bibitem[]{nos2003} P\'erez Mart\'{\i}nez, A., P\'erez Rojas,  H., \&  
 Mosquera Cuesta, H., 2003, Eur. Phys. J. C, 29, 111 

\bibitem[]{posiad91}Podsiadlowski, P., 1991, Nature 350, 136


---- 1995, MNRAS 274, 485  

---- 2000, ApJ 529, 946

\bibitem[]{Pons}Pons, J. A., et al., 2002, Phys. Rev. D 65, 104021


\bibitem[]{rezzolla2004} Rezzolla, L., and Akhmedov, B., 2004, MNRAS in press.


\bibitem[]{marsden01}Rothschild, R. E., Marsden, D. \& Lingenfelter, R. E.,
2001, report astro-ph/0105419.


\bibitem[]{shap-teuk}Shapiro, S. L. \& Teukolsky, S. A., 1983, {\it Black
Holes, White Dwarfs and Neutron Stars: The Physics of Compact Objects}
(Wiley \& Sons, New York)

\bibitem[]{suh00}Suh, I-S. \& Mathews, G. J., 2000, ApJ 530, 949.

 \bibitem[]{tauris00}Tauris, Th., Van den Heuvel, E. P. J. \& Savonije, G. J., 
2000, ApJ 530, L93.

 \bibitem[]{thompson}Thompson C. \& Duncan,  R. C., 1993, ApJ {408}, 194.

---- 1995, MNRAS {275}, 255

---- 1996, ApJ {473}, 322


\bibitem{paradijs}van Paradijs, J. et al., 1995a, ApJ 447, L33 and
1995b, A \& A 303, L25.

\bibitem[]{woods} Woods, P., et al., 1999, Ap. J. 527, L47


\bibitem[]{kouveliotou98-12}Xilouris, K., et al., 1998, ``Discovery of PSR
J1907+0918", AAS, {139.4106X}



\end{thebibliography}
\end{document}